\documentstyle[aps,epsf,preprint,prb]{revtex}
\tightenlines
\begin{document}

\bibliographystyle{unsrt}
\draft
%**end of header

\title{Comparison of Power Dependence of Microwave Surface 
Resistance of Unpatterned and Patterned YBCO Thin Film
}

\author{H. Xin,$^{\dag \Diamond *}$ D. E. Oates,$^{\dag
\Diamond}$
 A. C. Anderson,$^{\Diamond}$ R. L. Slattery,$^{\Diamond}$ G. Dresselhaus,$^{\dag *}$ 
 M.~S.~Dresselhaus$^{\dag}$}
\address{$^{\dag}$ Massachusetts Institute of Technology, Cambridge, MA 02139\\
$^{\Diamond}$ Lincoln Laboratory,  Lexington, MA 02420\\}
\address{$^*$AFRL, Hanscom AFB, Bedford, MA 01731\\}
\date{\today}
\pagebreak
\maketitle
\begin{abstract}

The effect of the patterning process on the
nonlinearity of the microwave surface resistance $R_S$ of YBCO thin
films is investigated. With the use of a sapphire dielectric resonator and a stripline
resonator, the microwave $R_S$ of YBCO thin films was measured before
and after the patterning process, as a function of temperature and the
rf peak magnetic field in the film. The microwave loss was also modeled,
assuming a $J_{rf}^2$ dependence of $Z_S(J_{rf})$ on current density $J_{rf}$. Experimental
and modeled results show that the patterning has no observable effect on the microwave residual
$R_S$ or on the power dependence of $R_S$.

\end{abstract}
%\pacs{}
\pagebreak
%\pacs{Ms number Bxxxx. PACS numbers: 74.25.Nf,74.50.+r,74.76.Bz}
%\narrowtext
%\begin{multicols}{2}[]

\section{Introduction}
\label{introduction}

The nonlinearity (power dependence) of the microwave (rf) surface
impedance $Z_S$ of the high-$T_c$ superconductors is important for both
practical applications and fundamental materials understanding.
Nonlinearity in the surface impedance not only degrades the power handling
ability of devices but also causes intermodulation and harmonic
generation problems at moderate power levels.  There has been a large
amount of effort to investigate the nonlinear $Z_S$,
but the origins of these phenomena are still not understood. Since most of the measurements have
been performed on photolithographically patterned stripline or
coplanar waveguide devices, one of the suggested explanations for the nonlinearity is
defects and damage caused by patterning. All
patterned devices have some current crowding near their edges. For
example, stripline and microstrip devices have current distributions
that peak sharply near their edges, where the patterning processes have a
relatively large influence. 

A SEM picture focusing on the edge of a patterned YBCO thin film is shown in Fig.~\ref{fig1}. 
The edge is not straight as can be seen in the picture and the
roughness of the edge is of the order of a superconducting penetration depth. This
edge roughness results from the initial imperfections of the
photoresist pattern and from nonuniformities of the etching
process. The edge roughness led to the speculation that the
condition of the YBCO line edge was influencing the power dependence,
and this was the initial motivation for this work. The bright
particles on the surface of the YBCO film are the well-known copper
oxide outgrowths that are present in all films whose stoichiometry is
not exact\cite{alfredo93}. We have found that outgrowths do not adversely affect the
microwave properties and therefore it is unnecessary to make large
efforts to deposit films with exactly the correct stoichiometry. We have also
observed that improvement of the surface morphology to make smoother
films than that shown in Fig.~\ref{fig1} does not improve their
microwave properties either with regard to their residual $R_S$ or their power dependence.

In the present work, we used both a sapphire dielectric resonator and
a stripline resonator to directly compare the power
dependence of the microwave surface resistance $R_S$ of the \it same
\normalsize YBCO thin films, first unpatterned in the dielectric
resonator and then patterned in the stripline resonator.  Experimental
procedures and results are shown in the following sections.  To help
understand our experimental results better, we have also modeled the
$R_S$ as measured in the dielectric resonator and in the stripline
resonator using the following simple phenomenological assumption, 
\begin{equation}
{\rho}(\vec{r}) = {\rho}_{0} + {\rho}_{2} {J_{rf}(\vec{r})}^2, 
\label{eq1.1}
\end{equation} 
\begin{equation}
{\lambda}(\vec{r}) = {\lambda}_{0} + {\lambda}_{2} {J_{rf}(\vec{r})}^2 
\label{eq1.2}
\end{equation}
where ${\rho}(\vec{r})$ and ${\lambda}(\vec{r})$ are, respectively,
the real part of the complex local 
resistivity and penetration depth, $J_{rf}(\vec{r})$ is the local current
density in the film, and ${\rho}_2$ and ${\lambda}_2$ are the corresponding
nonlinear coefficients.

%\nopagebreak
\section{Experimental Method}
\label{Method}

In order to study the effects of defects and damage introduced in the patterning
process on the power dependence of the microwave surface resistance of YBCO thin films, 
we measured the nonlinear microwave
frequency surface resistance at 10.7\,GHz of unpatterned,
2-inch-diameter, YBCO films using a specially designed sapphire
dielectric resonator. The unpatterned films were then patterned using
standard photolithography and wet chemical etching with 0.1\%
phosphoric acid to make stripline resonators with resonant frequencies
$f_0$ = $n$$f_1$, where $n = 1, 2$, ... and $f_1$ = 1.5\,GHz. The
patterning and dicing procedures are described in Fig.~\ref{fig2}. The microwave
surface resistance was then measured as a function of rf power for
the patterned striplines.  Comparing our results ``before'' and ``after'' the
patterning process, directly clarifies whether the patterning
contributes to the power dependence of the microwave surface resistance
observed for YBCO thin films.

%\nopagebreak
\subsection{Sapphire Dielectric Resonator}
\label{sdr}

We have designed a sapphire dielectric resonator shown in Fig.~\ref{fig3} to
measure the microwave surface resistance and especially the power
dependence of unpatterned YBCO thin films. The resonator incorporates
some unique features described below that allow high-power
measurements to be carried out without heating the sample via dissipated power. Figure~\ref{fig4} is a
cut-away view of the dielectric resonator.  The resonator was made from a
$\frac{\rm 1}{\rm 2}$-inch-diameter and $\frac{\rm 1}{\rm 4}$-inch-high cylindrical sapphire 
puck that is grounded by two 2-inch-diameter YBCO wafers (see left hand side of Fig.~\ref{fig2}). 
The resonator has fixed output coupling and in-situ-adjustable input
coupling to ensure critical input coupling (no reflection at the input end) for high-power
measurements. The resonator is packaged in an oxygen-free copper case,
as shown in Fig.~\ref{fig3}.  Special thermal contacts were designed to avoid
heating problems for high-power experiments.  A copper pressure plate
was employed on top of the upper YBCO wafer. Springs were used (as shown
in Fig.~\ref{fig3}) to ensure that the pressure on the YBCO is independent 
of temperature.  A gold-plated copper foil was soldered onto the copper pressure plate to enhance
thermal conduction to the copper case from the top YBCO film which was more likely to be
affected by heating than the bottom wafer that was anchored to the
copper base plate. Thermal-conducting grease was applied between the
sapphire puck and the two grounding wafers. The whole package 
was operated in a vacuum environment in a cryocooler with the bottom plate bolted to the cold 
finger with an indium foil for thermal contact. A temperature sensor and a 25-$\Omega$ 
heater were mounted on the copper case of the resonator with
indium foil to monitor and control the temperature.

The sapphire dielectric resonator was operated at the $TE_{011}$ mode for which
the center frequency was 10.7\,GHz. The loaded quality factor $Q_L$, reflection ($S_{11}$, $S_{22}$) 
and transmission ($S_{21}$) coefficients were measured as a function of
input power and temperature. The input coupling loop was tuned in-situ to maintain
critical coupling at each temperature and input power level. Critical
coupling on the input maximizes the circulating power and rf magnetic
field in the resonator. The output coupling loop was fixed and weakly
coupled during each measurement. The unloaded quality factor $Q_0$
was obtained from the measured $S$ parameters and $Q_L$, and the surface resistance
was deduced from $R_S$ = $G/{Q_0}$, where $G$ is the calculated geometric factor of the 
resonator.\cite{shen91} The results for $R_S$ are presented as a function of peak magnetic
field in the resonator. The $R_S$ measured by this method is the
average of $R_S$ in the top and bottom YBCO films.

The $Q_L$ was measured using a standard frequency-domain 3-dB bandwidth
method at a low input power where the $Q$ is independent of power. At
higher input power levels, a time-domain method was employed to reduce
heating effects.  In this method, a CW microwave pulse with enough
duration to fully charge the resonator was sent into the
resonator. After the input pulse is turned off, we can calculate $Q_L$ from 
the decay rate of the output signal that is proportional to $exp[-{\omega}t/{Q_L}]$. By
use of the time-domain method with this special thermal design, the
$R_S$ could be measured without heating problems for input powers up to 40\,dBm, which
corresponds to a rf peak magnetic field of $\geq$100\,Oe. 
Surface resistance measurements on two pairs of unpatterned YBCO films
were carried out for temperatures in the range 30 to 80\,K. 

In the analysis of the experimental data, we have, as is usual, treated the
losses of the YBCO superconducting ground planes as a perturbation
that does not affect the field or current distribution, while ignoring 
the losses due to the sapphire puck itself and to the
copper case. The assumption that only the YBCO films contribute to the
power loss is expected to be valid for the following
reasons. A good-quality sapphire crystal is almost lossless at the
cryogenic temperature where we operated, since the $\tan \delta$(loss tangent) is estimated\cite{shen91,blair89} to
be on the order of $10^{-9}$. For YBCO films, a $Q_0$ of
about $10^6$ is observed at low temperatures. A pair of 2-inch-diameter Nb films have 
yielded $Q_0$ values greater than $5\times 10^6$. That the Nb films gave a higher $Q$ than those 
of YBCO indicates that the loss in the sapphire puck is
negligible compared to that of the YBCO films. We estimated that the loss in
the copper case is very small and yields a $Q$ greater
than $2.5\times 10^8$, much higher than that of the YBCO films.

\nopagebreak
\subsection{Stripline Resonator}
\label{strip}
After the microwave surface resistance of the unpatterned YBCO films
was measured with the sapphire dielectric resonator, the films were
patterned and stripline resonators were made. A standard patterning
process with wet chemical etching was employed. As shown in
Fig.~\ref{fig2}, each 2-inch wafer was made into four striplines and
four ground planes.  One piece from each of the top and the bottom YBCO
wafers used with the dielectric resonator in Fig.~\ref{fig3} was measured.  
A description of the
stripline resonator used here can be found in detail elsewhere.\cite{oates90,oates92} 
Similar to the dielectric
resonator method, the loaded quality factor $Q_L$ was measured for the stripline as a
function of microwave input power up to 35\,dBm, corresponding to a rf
peak magnetic field of 1000\,Oe in the resonator. One advantage of the
stripline resonator is that the $R_S$ at different overtone frequencies
can be easily measured. By measuring overtones, we could measure the
$Q$ at frequencies close to that of the dielectric resonator even though the fundamental 
resonance of the stripline resonator is 1.5\,GHz.

\nopagebreak
\section{Experimental Results}
\label{results}
The $R_S$ as a function of rf peak magnetic field $H_{\rm max}$ at different temperatures was measured
for two pairs of unpatterned YBCO films, denoted as pair 1 and 2 (see
Fig.~\ref{fig3}). The results are shown in Fig.~\ref{fig5}. For pair 1, no
nonlinearity in $R_S$ of the unpatterned wafers was observed up to
80\,K even at +40\,dBm, the maximum available power,
and corresponding to $H_{rf}$ values up to 200\,Oe.
As mentioned above, $R_S$ obtained for the unpatterned wafers is an average surface resistance of the
pair of films. Figure~\ref{fig6}(a) compares $R_S$ of the unpatterned and
patterned films at 35\,K for pair 1. The open circles represent the $R_S$ of the patterned
film from the top wafer measured in the dielectric resonator, and the
crosses denote the $R_S$ of the patterned film from the bottom wafer. The
average of those two curves (dashed line) agrees with the $R_S$ of the
unpatterned wafers very well and suggests that there is no
nonlinearity for $H_{\rm max}$ up to 500\,Oe. Figure~\ref{fig6}(b) shows the results at
75\,K for pair 1. For the case of stripline resonators, because of package modes
interfering with some of the overtone modes, the $R_S$ could not be measured at the mode closest 
in frequency to that of the dielectric resonator. Therefore $R_S$ was measured at
the closest frequencies possible.  The results for $R_S$ were then scaled
assuming $R_S$ $\propto$ $\omega^2$. For pair 1 the $R_S$ of the
unpatterned films was measured at 10.7\,GHz, and the $R_S$ of the two
patterned films was measured at 12\,GHz and 7.5\,GHz.
  
From the results in Fig.~\ref{fig6}, we conclude that for pair 1, no degradation in the residual
surface resistance was observed in the patterned stripline films up to a $H_{\rm max}$ of $\sim$ 200\,Oe. 
No power dependence was observed up to the maximum input power available for the unpatterned 
films, and no effects of patterning on the power handling of YBCO
films could be found either at 35\,K or at 75\,K for pair 1.

We grew another pair of 2-inch-diameter YBCO wafers pair 2, under a different deposition temperature 
with the intention to decrease the power handling of the microwave surface resistance.
For pair 2 at 35\,K no power dependence of the $R_S$ was observed in
the dielectric resonator up to the maximum input power, 
similar to the results obtained for pair 1.  However, at $T =$ 75\,K, 
a nonlinearity in the $R_S$ was observed above $H_{\rm max}$ $\sim$ 100\,Oe. 
Results for the patterned films are shown together with that
of the unpatterned films in Fig.~\ref{fig7}. In this case, $R_S$ was measured at 7.5
GHz for both patterned films in the stripline resonator. The measured residual
surface resistance of the unpatterned and patterned films
agrees to within 10$\%$. The $R_S(H_{\rm max})$ for the unpatterned films actually
turns up at a somewhat lower $H_{\rm max}$ than for the patterned films in the stripline resonator.
The apparently better power handling of the stripline resonator can be explained using the same
surface impedance by a simple model introduced below. The $H_{\rm
max}$ is the rf peak magnetic field 
in the two types of resonators, but the two types of resonators have very different field distributions. 
In the next section, we use the different current/field profiles for both the sapphire dielectric and 
stripline resonators to model the measured power dependence of the unpatterned and patterned films.

\section{Model}
\label{model}
The current profiles in the films for the dielectric and stripline
resonators are different, as shown in Fig.~\ref{fig8}. For the
dielectric resonator with a film diameter much larger
than the diameter of the sapphire puck so that the edge effects can be ignored,
the current density $\vec{J}$ is in the $\phi$ direction and has a dependence on radius
given by\cite{shen91,kajfez86}
\begin{equation} 
J(r) = {\frac{H(r)}{\lambda_0}} = \left\{ \begin{array}{ll}
                                              A{\frac{\beta}{{\xi_1}{\lambda_0}}}{J_1}(\xi_{1}r) & \mbox{for $0<r<a$} \\
                                      A{\frac{\beta}{{\lambda_0}{\xi_2}}}{\frac{{J_0}({\xi_1}a)}{{K_0}({\xi_2}a)}}{K_1}({\xi_2}r) & \mbox{for $a<r<R$}
\end{array}
\right.  
\label{eq1.3}
\end{equation} 
where $J_0$, $J_1$, $K_0$, $K_1$ are various Bessel functions,
$\beta$, $\xi_1$ and $\xi_2$ are constants determined by the geometry
of the resonator, \it a \normalsize is the radius of the sapphire puck and
$R$ is the radius of the superconducting films. The current density thus peaks near the center of 
the film as shown in Fig.~\ref{fig8}.  For the stripline resonator, the current density
$\vec{J}$ has been calculated numerically in Ref\onlinecite{sheen91}. For the purpose of simplicity in the
calculation, we approximated $J(x)$ with the following analytical form,\cite{duzer81} 
\begin{equation} J(x) = \left\{ \begin{array}{ll}
{J_s(0)}{[1-{(\frac{2x}{w})}^2]}^{-\frac{1}{2}} & x\ll w\\
{J_s(0)}(\frac{1.165}{\lambda_0})(wb)^{\frac{1}{2}}\exp(-\frac{(w/2-|x|)b}{{\lambda_0}^2})
& x\approx w \end{array} \right. 
\label{eq1.4} 
\end{equation} 
where, $w$ and $b$ are, respectively, the width and thickness of the
film. This analytical approximation agrees with the numerically
calculated current distribution to within 5$\%$ in our case.
Thus $J(x)$ 
peaks sharply at the edges of the patterned film for the stripline resonator. For the dielectric
resonator, as is obvious in Eq.~(\ref{eq1.3}), $J(x)$ is related directly
to the rf peak magnetic field $H_{\rm max}$. For the stripline resonator, $H_{\rm max}$ can be calculated numerically from
the current distribution in the stripline.\cite{sheen91} For the same 
maximum field $H_{\rm max}$ in the dielectric and stripline resonators, the portion
of the film carrying a high current density is larger for the dielectric
resonator due to the broader peak in the $J(r)$ distribution of the
dielectric resonator. Therefore an apparently better power handling in
the stripline resonator is not surprising. 

The quality factors $Q_0$($H_{\rm max}$) and in turn $R_S$($H_{\rm max}$) of both resonators 
have been modeled as a function of rf peak magnetic field using the current distributions
given above, with the assumption for the impedance is given by Eqs.~(\ref{eq1.1}) and
(\ref{eq1.2}), and
\begin{equation}
 {Q_0} = \omega\frac{W_{\rm stored}}{P_{\rm diss}},
\label{eq1.5}
\end{equation}
where $W_{\rm stored}$ is the total energy stored and
$P_{\rm diss}$ is the power dissipated in the resonator.  The parameters
$\rho_0$ and $\lambda_0$ were taken from experimentally measured
values, $\rho_0 = 7.2 \times 10^{-11}$\,$\Omega$\rm {m}, $\lambda_0$ = 0.2\,$\mu$\rm {m}, that fit
the low-field, linear part of $R_S$ and the temperature dependence of the 
resonant frequency. The nonlinear parameters of
resistivity and penetration depth used in the calculations in the superconducting phase
were $\rho_2 = 1.0 \times 10^{-35}$\,$\Omega$$\rm {m^{5}/A^{2}}$ and $\lambda_2 = 2.5 \times 10^{-29}$\,$\rm {m^{5}/A^{2}}$. 
These parameters were taken to fit the $R_S$ obtained from the dielectric resonator
measurements. The unloaded quality factor for the stripline resonator was then
calculated with the same set of parameters. At low rf field, calculated values of $Q_0$
of $7\times 10^5$ and 6000, respectively, were obtained for the dielectric and stripline
resonators, and these values are consistent with the measured values. The modeled results also
show that the power dependence of $R_S$ for both the dielectric and stripline resonators is 
quantitatively consistent with the measured power dependence for both resonators. The modeled $R_S$ 
of the dielectric resonator turns up at a smaller value of $H_{\rm
max}$ than that of the stripline 
resonator, which agrees with the experimental result well as shown in Fig.~\ref{fig9}.
 
\section{Conclusions and Discussion}
\label{conclusion}
We have measured the microwave surface resistance of the \it same
\normalsize YBCO thin films, before and after patterning, using a
sapphire dielectric resonator and a stripline resonator. 
A phenomenological model of the $R_S$$(H_{\rm max})$ in the
two cases is presented. Based on this model, the calculated results fit the
experimental data well in both cases using the same materials
parameters. Therefore, we conclude that our results are consistent with
no damage or degradation of the power dependence of the microwave surface resistance
due to the patterning of the films.  The procedure described in
this paper can also be used to test other patterning processes such as
ion beam etching, etc.

\acknowledgments

This work was supported by the Air Force Office of Scientific
Research. The authors express their gratitude to Bob Koneizcka and Earle Macedo for applying their
tireless efforts and effective skills to the preparation of the devices
used in this study, and to Peter Murphy for providing the SEM pictures used in
this study.

%\bibliography{/usr2/mgm/mgm-bib/mgm,/usr2/mgm/mgm-bib/hitc,/home3/hao/research/hitc}
%\bibliography{/home3/hao/research/hitc}

\narrowtext

\begin{figure}
%\begin{narrowtext}
%\[
%\epsfxsize 10.5cm
%\centerline{\epsffile{/home3/hao/research/pattern/fig1.ps}}
%\]
\caption{A SEM picture of the edge of a patterned YBCO stripline.}
\label{fig1}
\end{figure}

\begin{figure}
%\[
%\epsfxsize 10.5cm
%\centerline{\epsffile{/home3/hao/research/newfig1/fig2.eps}}
%\]
\caption{YBCO film before(left) and after(right) the patterning and dicing processes.}
\label{fig2}
\end{figure}

\begin{figure}
%\[
%\epsfxsize 10.5cm
%\centerline{\epsffile{/home3/hao/research/newfig/nfig3.ps}}
%\]
\caption{Structure of the dielectric resonator.}
\label{fig3}
\end{figure}

\begin{figure}
%\[
%\epsfxsize 10.5cm
%\centerline{\epsffile{/home3/hao/research/newfig/nfig4.ps}}
%\]
\caption{Cut-away view of the dielectric resonator.}
\label{fig4}
\end{figure}

\begin{figure}
%\[
%\epsfxsize 10.5cm
%\centerline{\epsffile{/home3/hao/research/newfig/nfig5.ps}}
%\]
\caption{$R_S$ vs. rf peak magnetic field($H_{max}$) for the unpatterned films (pair 1) for various temperatures.}
\label{fig5}
\end{figure}

\begin{figure}
%\[
%\epsfxsize 10.5cm
%\centerline{\epsffile{/home3/hao/research/newfig/nfig6a.ps}}
%\]
%\[
%\epsfxsize 10.5cm
%\centerline{\epsffile{/home3/hao/research/newfig/nfig6b.ps}}
%\]
\caption{ $R_S$ vs. rf peak magnetic field($H_{max}$) for both unpatterned and patterned
films (pair 1) at (a) 35\,K and (b) 75\,K. The solid circles represent data of the unpatterned films; the open circles and crosses
represent data of the patterned films from the top and bottom wafers, respectively; the dashed line represents
the average $R_S$ of patterned films from the top and bottom wafers.}
\label{fig6}
\end{figure}

\begin{figure}
%\[
%\epsfxsize 10.5cm
%\centerline{\epsffile{/home3/hao/research/newfig/nfig7.ps}}
%\]
\caption{$R_S$ vs. rf peak magnetic field for both unpatterned wafers 1 and 2 and the patterned
films (pair 2) at 75\,K. The squares represent data of the unpatterned films; the circles and the triangles represent 
data of the patterned films from the top and bottom wafers, respectively.}
\label{fig7}
\end{figure}

\begin{figure}
%\[
%\epsfxsize 10.5cm
%\centerline{\epsffile{/home3/hao/research/fig8.ps}}
%\]
\caption{Current distributions of both the stripline resonator (left) and the dielectric (sapphire) resonator (right).}
\label{fig8}
\end{figure}

\begin{figure}
%\[
%\epsfxsize 10.5cm
%\centerline{\epsffile{/home3/hao/research/newfig/nfig9.ps}}
%\]
\caption{Modeled and measured $R_S$ vs. rf peak magnetic field for both the dielectric resonator and the stripline resonator. 
The triangles represent the measured $R_S$ for the dielectric resonator; the circles represent the measured
$R_S$ for the stripline resonator; the dashed line represents the modeled $R_S$ for the dielectric resonator;
the solid line represents the modeled $R_S$ for the stripline resonator.}
\label{fig9}
\end{figure}

%\end{multicols}
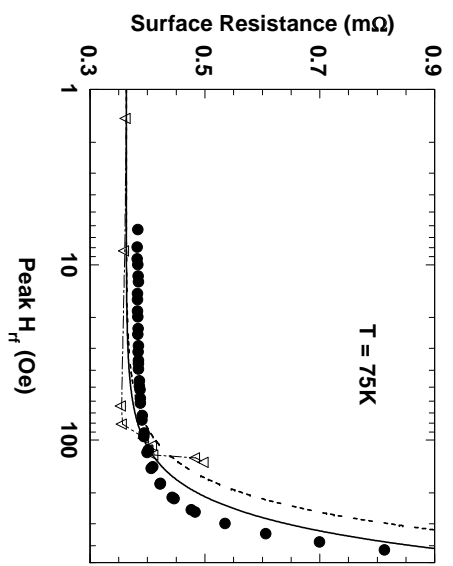
\end{document}